\documentclass[showpacs,preprintnumbers,amsmath,amssymb,twocolumn,superscriptaddress,prl]{revtex4}
\usepackage{graphicx}
\usepackage{amsfonts}
\usepackage{amsmath, amsthm, amssymb}

\newcommand{\Imag}{{\Im\mathrm{m}}}   
\newcommand{\im}{\mathrm{i}}        
\newcommand{\ve}[1]{\mathbf{#1}}
\DeclareMathOperator{\diag}{diag} 
\newcommand{\x}{\lambda}  
\newcommand{\y}{\rho}     
\newcommand{\vk}{\ve{k}} 
\newcommand{\vp}{\ve{p}} 
\newcommand{\vq}{\ve{q}} 
\newcommand{\ca}[2][]{c_{#2}^{\vphantom{\dagger}#1}} 
\newcommand{\cc}[2][]{c_{#2}^{{\dagger}#1}}          
\newcommand{\da}[2][]{d_{#2}^{\vphantom{\dagger}#1}} 
\newcommand{\dc}[2][]{d_{#2}^{{\dagger}#1}}          
\newcommand{\ga}[2][]{\gamma_{#2}^{\vphantom{\dagger}#1}} 
\newcommand{\gc}[2][]{\gamma_{#2}^{{\dagger}#1}}          
\newcommand{\Tkp}[1]{T_{\vk\vp#1}}  

\newcommand{\e}[1]{\mathrm{e}^{#1}}
\newcommand{\dif}{\mathrm{d}} 
\newcommand{\mean}[1]{\langle#1\rangle}
\newcommand{\abs}[1]{|#1|}

\newcommand{\pauli}[1][\alpha\beta]{\boldsymbol{\sigma}_{#1}^{\vphantom{\dagger}}}
\newcommand{\wdf}[2][\frac{1}{2}]{{\cal{D}}_{#2}(\vartheta)} 
\newcommand{\wdfs}[2][\frac{1}{2}]{{\cal{D}}^{\;2}_{#2}(\vartheta)} 

\newcommand{\eq}{Eq.}
\newcommand{\ie}{\textit{i.e. }}
\newcommand{\eg}{\textit{e.g. }}
\newcommand{\etal}{\emph{et al.}}

\def\i{\mathrm{i}}

\begin{document}
\title[Interplay between ferromagnetism and superconductivity in
tunneling currents]{Interplay between ferromagnetism and
  superconductivity in tunneling currents}
\author{M. S. Gr{\o}nsleth}
\affiliation{Department of Physics, Norwegian University of
Science and Technology, N-7491 Trondheim, Norway}
\author{J. Linder}
\affiliation{Department of Physics, Norwegian University of
Science and Technology, N-7491 Trondheim, Norway}
\author{J.-M. B{\o}rven}
\affiliation{Department of Physics, Norwegian University of
Science and Technology, N-7491 Trondheim, Norway}
\author{A. Sudb{\o}}
\affiliation{Department of Physics, Norwegian University of
Science and Technology, N-7491 Trondheim, Norway}
\date{Received \today}
\begin{abstract}
  We study tunneling currents in a model consisting of two non-unitary
  ferromagnetic spin-triplet superconductors separated by a thin
  insulating layer. We find a novel interplay between ferromagnetism
  and superconductivity, manifested in the Josephson effect. This offers the
  possibility of tuning dissipationless currents of charge and spin in
  a well-defined manner by adjusting the magnetization direction on
  either side of the junction.
\end{abstract}
\maketitle
The coexistence of ferromagnetism (FM) and superconductivity (SC) has
recently been experimentally confirmed \cite{saxena,pfleiderer,aoki}.
This offers the possibility of observing new and interesting physical
effects in transport of spin and charge. The spin-singlet character of
Cooper pairs in conventional superconductors suggests that FM and SC
are mutually excluding properties for a material. Indeed, coexistence
of magnetic order with singlet superconductivity is not possible for
uniform order parameters \cite{kulic} but can occur if the
superconductor is in a Fulde-Ferrell-Larkin-Ovchinnikov (FFLO)
\cite{fflo} state with modulated ferromagnetic and superconducting
order parameters characterized by a non-zero wave vector $\mathbf{q}$.
On the other hand, spin-triplet Cooper pairs \cite{maeno,ishida} are
in principle perfectly compatible with ferromagnetism. For instance,
odd-in-frequency $S_z=0$ spin-triplet superconductivity in
superconductor-ferromagnet structures has been much studied in the
literature \cite{ber}. The synthesis of superconductors exhibiting
ferromagnetism, with simultaneously broken U(1) and O(3) symmetries,
are of considerable interest from a fundamental physics point of view,
and moreover opens up a vista to a plethora of novel applications.
This has been the subject of theoretical research in {\it e.g.}  Refs.
\cite{brataas1}, and has a broad range of possible applications. Also,
focus on hybrid systems of ferromagnets and superconductors has arisen
from the aspiration of utilizing the spin of the electron as a binary
variable in device applications. This has led to spin current induced
magnetization switching \cite{kiselev}, and suggestions have been made
for devices such as spin-torque transistors \cite{brataas2} and
spin-batteries \cite{brataas3}. Moreover, spin supercurrents have a
long tradition in $^3$He \cite{leggett1975}, while recent work has
focused on dissipationless spin-currents in unitary spin-triplet
superconductors \cite{asanov}.
\par
\begin{figure}[h!]
  \centering
  \resizebox{0.48\textwidth}{!}{
    \includegraphics{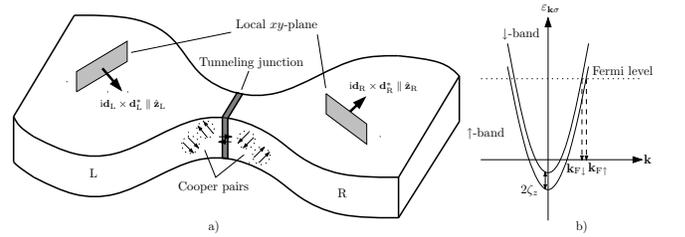}}
  \caption{a) Tunneling of Cooper pairs between two non-unitary FMSC with
    non-collinear magnetization. b) Band-splitting for
    $\uparrow,\downarrow$ electrons in the presence of a magnetization
    in $\hat{\mathbf{z}}$-direction, leading to a suppression of interband-pairing.}
  \label{fig:junction}
\end{figure}

This Letter addresses the case of two $p$-wave superconductors arising
out of a ferromagnetic metallic state, separated by a tunneling
junction. Such states have been suggested to exist on experimental
grounds, in compounds such as RuSr$_2$GdCu$_2$O$_8$ \cite{tallon1999},
UGe$_2$ \cite{saxena}, ZrZn$_2$ \cite{pfleiderer}, and URhGe
\cite{aoki}, and have been studied theoretically in \eg Refs.
\cite{kirkpatrick, machida, bedell}. Coexisting FM and spin-triplet SC
have also been proposed to arise out of half-metallic ferromagnetic
materials such as CrO$_2$, and the alloys UNiSn and NiMnSb
\cite{rudd_pickett_1998}. We compute the Josephson contribution to the
tunneling currents, both in the charge- and spin-channel, within
linear response theory using the Kubo formula. Our assumption is that
the superconducting order is that of spin-triplet pairing, and we
consider the analog of the A2-phase in $^3$He, \ie SC order parameters
that satisfy
$|\Delta_{\vk\uparrow\uparrow}|\neq|\Delta_{\vk\downarrow\downarrow}|
\neq 0$ and $\Delta_{\vk\uparrow\downarrow}=0$. In terms of the
$\mathbf{d}_\vk$-vector formalism \cite{leggett1975}, we then have a
non-unitary state since the average spin $\langle\mathbf{S}_\vk\rangle
= \i\mathbf{d}_\vk\times\mathbf{d}_\vk^\ast$ =
$\frac{1}{2}(|\Delta_{\vk\uparrow\uparrow}|^2 -
|\Delta_{\vk\downarrow\downarrow}|^2)\hat{\mathbf{z}}$ of the Cooper
pairs is nonzero. Such a scenario is compatible with uniform FM and SC
since the electrons responsible for ferromagnetism below the Curie
temperature $T_M$ condense into Cooper pairs with magnetic moments
aligned with the magnetization below the critical temperature $T_c$.
The choice of such a non-unitary state is motivated by the fact that
there is strong reason to believe that the correct pairing symmetries
in the ferromagnetic superconductors (FMSC) discovered so far are
non-unitary \cite{huxley,samokhin,machida}. The exchange field will
also give rise to a Zeeman-splitting between the $\uparrow,\downarrow$
conduction bands, thus suppressing the SC order parameter
$\Delta_{\vk\uparrow\downarrow}$ \cite{aoki}, as illustrated in Fig.
\ref{fig:junction}b).
\par
Another important issue to address is whether the SC and FM order
parameters coexist uniformly, or if they are phase-separated. One
possibility is that a spontaneously formed vortex lattice due to the
internal magnetization $\mathbf{m}$ is realized in a spin-triplet FMSC
\cite{tewari}. However, there have also been reports of uniform
superconducting phases in spin-triplet FMSC \cite{shopova}. A key
variable determining whether a vortex lattice appears or not is the
strength of the internal magnetization $\mathbf{m}$ \cite{mineev2}.
Current experimental data on ZrZn$_2$ and URhGe apparently do not
settle this issue unambiguously, while uniform coexistence of FM and
SC appear to have been experimentally verified in UGe$_2$
\cite{kotegawa}. Furthermore, a bulk Meissner state in the FMSC
RuSr$_2$GdCu$_2$O$_8$ has been reported in Ref. \cite{bernhard},
indicating the existence of uniform FM and SC as a bulk effect. Consequently, we will use bulk values for the order parameters and assume that they coexist uniformly. We emphasize that one in general should
take into account the possible suppression of the SC order parameter
in the vicinity of the tunneling interface due to the formation of
midgap surface states \cite{hu} which occur for certain orientations
of the SC gap. The pair-breaking effect of these states in
unconventional superconductors has been studied in \eg Ref.
\cite{tanuma}. A sizeable formation of such states
would suppress the Josephson current, although it is nonvanishing in
the general case.  Also, we use
bulk uniform magnetic order parameters, as in Ref. \cite{nogueira}.
The latter is justified on the grounds that a ferromagnet with
a planar order parameter is mathematically isomorphic to an $s$-wave
superconductor, where the use of bulk values for the order
parameter right up to the interface is a good approximation
due to the lack of midgap surface states. 
Moreover, we consider thin film FMSC such that the Lorentz-force
acting on the electrons will be unable to accelerate particles in a
direction parallel to the junction. Our model is illustrated in Fig.
\ref{fig:junction}a).
\par
The main result of this Letter is that the Josephson current in spin-
and charge-sector between two non-unitary FMSC can be controlled by
adjusting the relative magnetization orientation on each side of the
junction provided that spin-triplet Cooper pairs are present. Our
system consists of two FMSC separated by an insulating layer such that
the total Hamiltonian can be written as 
$H = H_\text{L} + H_\text{R} + H_\text{T}$, where L and R represents
the individual FMSC on each side of the tunneling junction, and
$H_\text{T}$ describes tunneling of particles through the insulating
layer separating the two pieces of bulk material. The FMSC Hamiltonian
is given by \cite{bedell} $H_\text{FMSC} = H_0 +
\sum_{\vk}\psi_{\vk}^\dag \cal{A}_{\vk}\psi_{\vk}$, where $H_0 =
JN\gamma(0)\mathbf{m}^2 +
\frac{1}{2}\sum_{\vk\sigma}\varepsilon_{\vk\sigma} +
\sum_{\vk\alpha\beta} \Delta_{\vk\alpha\beta}^\dag
b_{\vk\alpha\beta}$. Here, $\vk$ is the electron momentum,
$\varepsilon_{\vk\sigma} = \varepsilon_{\vk} - \sigma\zeta_z$,
$\sigma=\uparrow,\downarrow=\pm 1$, $J$ is a spin coupling constant,
$\gamma(0)$ is the number of nearest lattice neighbors,
$\mathbf{m}=\{m_x,m_y,m_z\}$ is the magnetization vector, while
$\Delta_{\vk\alpha\beta}$ is the superconducting order parameter and
$b_{\vk\alpha\beta} = \langle c_{-\vk\beta}c_{\vk\alpha}\rangle$ is
the two-particle operator expectation value. The ferromagnetic order
parameter is defined by $\zeta = 2J\gamma(0)(m_x-\i m_y)$ and $\zeta_z
= 2 J\gamma(0)m_z$.  We express the Hamiltonian in the basis
$\psi_{\vk} = (c_{\vk\uparrow}~c_{\vk\downarrow}~c_{-\vk\uparrow}^\dag
~c_{\mathbf{-k}\downarrow}^\dag)^{\text{T}}$, where $c_{\vk\sigma}$
($c_{\vk\sigma}^\dag$) are annihilation (creation) fermion operators.
Note that there is no spin-orbit coupling in our model, \ie inversion
symmetry is not broken. Consider now the 4$\times$4 matrix
\begin{equation}\label{eq:A}
{\cal{A}}_\vk = -\frac{1}{2} 
\begin{pmatrix}
-\varepsilon_\vk\mathbf{1} + \boldsymbol{\sigma}\cdot\boldsymbol{\zeta} & \i\mathbf{d}_\vk\cdot\boldsymbol{\sigma}\sigma_y\\
(\i\mathbf{d}_\vk\cdot\boldsymbol{\sigma}\sigma_y)^\dag & (\varepsilon_\vk\mathbf{1} - \boldsymbol{\sigma}\cdot\boldsymbol{\zeta})^\text{T}\\
\end{pmatrix}
\end{equation}
\noindent which is valid for a FMSC with arbitrary magnetization. As explained
in the introduction, we will study in detail a non-unitary equal-spin
pairing (ESP) FMSC as illustrated in Fig. \ref{fig:junction}a), \ie
$\Delta_{\vk\uparrow\downarrow}=\Delta_{\vk\downarrow\uparrow}=0$,
$\zeta= 0$ in Eq. (\ref{eq:A}). Since the quantization axes of the two
FMSC are not aligned, one needs to include the Wigner $d$-function
\cite{wigner} denoted by $\wdf{\sigma'\sigma}$ with $j=1/2$ to account
for the fact that a $\uparrow$ spin on one side of the junction is not
the same as a $\uparrow$ spin on the other side of the junction. The
spin quantization axes are taken along the direction of the
magnetization on each side, so that the angle $\vartheta$ is defined
by $\mathbf{m}_\text{R}\cdot\mathbf{m}_\text{L} =
m_\text{R}m_\text{L}\cos(\vartheta)$ where $m_i = |\mathbf{m}_i|$.
The tunneling Hamiltonian then reads $H_\mathrm{T} =
\sum_{\vk\vp\sigma\sigma'}\wdf{\sigma'\sigma}(
T_{\vk\vp}^{\vphantom{\dagger}}\cc{\vk\sigma}\da{\vp\sigma'} +
T_{\vk\vp}^{\vphantom{\dagger}\ast}\dc{\vp\sigma'}\ca{\vk\sigma}),$
where we neglect the possibility of spin-flips in the tunneling
process. The validity of the tunneling Hamiltonian approach requires
that the applied voltage across the junction is small. Here, we will
be concerned with the case of zero bias voltage, so that the tunneling
Hamiltonian approach is appropriate. Note that we distinguish between
fermion operators on each side of the junction corresponding to
$c_{\vk\sigma}$ and $d_{\vk\sigma}$.  Furthermore, we write the
superconducting order parameters as $ \Delta_{\vk\sigma\sigma} =
|\Delta_{\vk\sigma\sigma}|\e{\i(\theta_\vk +
  \theta^\text{R}_{\sigma\sigma})}$, where R (L) denotes the bulk
superconducting phase on the right (left) side of the junction while
$\theta_\vk$ is a general (complex) internal phase factor originating
from the specific form of the gap in $\vk$-space that ensures odd
symmetry under inversion of momentum, \ie $\theta_\vk = \theta_{-\vk}
+ \pi$.
\par 
For our system, the Hamiltonian takes the form $H_\text{FMSC} = H_0 +
H_A$, $H_A = \sum_{\vk\sigma}\phi_{\vk\sigma}^\dag A_{\vk\sigma}
\phi_{\vk\sigma}$, where we have chosen a convenient basis
$\phi_{\vk\sigma}^\dag = (c_{\vk\sigma}^\dag, c_{-\vk\sigma})$ that
block-diagonalizes ${\cal{A}}_\vk$, and defined $A_{\vk\sigma} =
\frac{1}{4}[2\varepsilon_{\vk\sigma}\sigma_z +
\Delta_{\vk\sigma\sigma}(\sigma_x+\i\sigma_y) +
\Delta_{\vk\sigma\sigma}^\dag(\sigma_x-\i\sigma_y)]$ with Pauli
matrices $\sigma_i$.
This Hamiltonian is diagonalized by a $2\times 2$ spin generalized
unitary matrix $U_{\vk\sigma}$, so that the superconducting sector is
expressed in the diagonal basis $\tilde{\phi}_{\vk\sigma}^\dagger =
\phi_{\vk\sigma}^\dagger U_{\vk\sigma} \equiv (\gc{\vk\sigma},
\ga{-\vk\sigma})$. Thus, $H_A =
\sum_{\vk\sigma}\tilde{\phi}_{\vk\sigma}^\dagger \tilde{A}_{\vk\sigma}
\tilde{\phi}_{\vk\sigma}$, in which $\tilde{A}_{\vk\sigma} =
U_{\vk\sigma}A_{\vk\sigma}U_{\vk\sigma}^{-1} =
\diag(\widetilde{E}_{\vk\sigma}, -\widetilde{E}_{\vk\sigma})/2$, and
$\widetilde{E}_{\vk\sigma} =
\sqrt{\varepsilon_{\vk\sigma}^2+\abs{\Delta_{\vk\sigma\sigma}}^2}$.

In order to find the spin and charge currents over the junction,
consider first the generalized number operator $N_{\alpha\beta}=
\sum_\vk\cc{\vk\alpha}\ca{\vk\beta}$. The transport operator in the
interaction picture then reads
$\dot{N}_{\alpha\beta}(t)=-\im\sum_{\vk\vp\sigma}[
\wdf{\sigma\beta}T_{\vk\vp}\cc{\vk\alpha}(t)\da{\vp\sigma}(t) \e{-\im
  teV}
-\wdf{\sigma\alpha}T_{\vk\vp}^\ast\dc{\vp\sigma}(t)\ca{\vk\beta}(t)
\e{\im teV}]$,
where $eV\equiv \mu_\text{L}-\mu_\text{R}$ is an externally applied
potential. The general current across the junction can be written
\begin{equation}\label{eq:currents}
  \mathbf{I}(t) = 
  \sum_{\alpha\beta}\boldsymbol{\tau}_{\alpha\beta}\langle 
  \dot{N}_{\alpha\beta}(t)\rangle,\; 
  \boldsymbol{\tau} = (-e\boldsymbol{1},\boldsymbol{\sigma})
\end{equation}
such that the charge-current is $I^\text{C}(t) = I_0(t)$ while the
spin-current reads $\mathbf{I}^\text{S}(t) = (I_1(t),I_2(t),I_3(t))$.
Note that Eq. (\ref{eq:currents}) contains both the single-particle
(sp) and two-particle (tp) contribution. The concept of a spin-current
in this context refers to the rate at which the spin-vector
$\mathbf{S}$ on one side of the junction changes \textit{as a result
  of tunneling across the junction}, \ie
$\dot{\mathbf{S}}=\i[H_\mathrm{T},\mathbf{S}].$ As there is no
spin-orbit coupling in our system, this definition of the spin-current
serves well \cite{shi2}. The spatial components of
$\mathbf{I}^\text{S}$ are defined with respect to the corresponding
quantization axis.  We compute the currents by the Kubo formula,
$\mean{\dot{N}_{\alpha\beta}(t)} = -\im\int_{-\infty}^{t}\dif t'
\mean{[\dot{N}_{\alpha\beta}(t),H_\mathrm{T}(t')]}$, where the right
hand side is the statistical expectation value in the unperturbed
quantum state, \ie when the two subsystems are not coupled. We now
focus on the two-particle charge-current and
$\hat{\mathbf{z}}$-component of the spin-current, such that only
$\alpha=\beta$ contributes in Eq. (\ref{eq:currents}).
\par 
Using linear response theory with the Matsubara formalism, one arrives
at $\mean{\dot{N}_{\alpha\alpha}(t)}_\mathrm{tp} =
2\sum_\sigma\Im\text{m}\{ \Psi_{\alpha\sigma}(eV) \e{-2\im teV} \},$
where $\Psi_{\alpha\sigma}(eV)$ is obtained by performing analytical
continuation $\i\omega_n\to eV+\i 0^+$ ($\omega_n = 2\pi n
k_\text{B}T, n = 1,2,3\ldots$) on
$\widetilde{\Psi}_{\alpha\sigma}(\i\omega_n) = -\int^{1/k_\text{B}T}_0
\text{d}\tau \e{\i\omega_n\tau} \langle\mathrm{T}_\tau
M_{\alpha\sigma}(\tau)M_{\alpha\sigma}(0)\rangle,$ where
$M_{\alpha\sigma}(t) = \sum_{\vk\vp}\wdf{\sigma\alpha}T_{\vk\vp}
\cc{\vk\alpha}(t)\da{\vp\sigma}(t),$ while $T$ is the temperature, and
$\text{T}_\tau$ denotes the time-ordering operator. Explicitly, we
find
\begin{equation}
  \label{eq:twopartlambda}
  \Psi_{\alpha\sigma}(eV)=
  -\sum_{\substack{\vk\vp\\\x,\y=\pm}}
  \wdfs{\sigma\alpha}\abs{T_{\vk\vp}}^2
  \frac{\Delta_{\vk\alpha\alpha}^\ast\Delta_{\vp\sigma\sigma}}{4E_{\vk\alpha}E_{\vp\sigma}}\Lambda_{\vk\vp\alpha\sigma}^{\x\y}(eV)
\end{equation}
where $E_{\vk\sigma} = \sqrt{\xi_{\vk\sigma}^2 +
  |\Delta_{\vk\sigma\sigma}|^2}$, $\xi_{\vk\sigma} =
\varepsilon_{\vk\sigma} - \mu_\text{R}$ (R $\to$ L for $\vk\to\vp$),
and $\Lambda_{\vk\vp\alpha\sigma}^{\x\y}(eV)$ = $\lim_{\i\omega_n\to
  eV+\i 0^+}$$ \x[f(E_{\vk\alpha}) - f(\x\y
E_{\vp\sigma})]/[\im\omega_\nu + \y E_{\vk\alpha}-\x
E_{\vp\sigma}];\quad \x,\y=\pm 1.$

In Eq. (\ref{eq:twopartlambda}), we have used that
$T_{-\vk,-\vp}=T_{\vk\vp}^\ast$ which follows from time reversal
symmetry, and $f(x)$ is the Fermi distribution. Note that the chemical
potential has been included in the excitation energies
$E_{\vk\alpha}$. In general, Eq. (\ref{eq:twopartlambda}) will give
rise to a term proportional to
$\cos(\theta^\text{L}_{\sigma\sigma}-\theta^\text{R}_{\alpha\alpha})$,
the quasiparticle interference term, in addition to
$\sin(\theta^\text{L}_{\sigma\sigma}-\theta^\text{R}_{\alpha\alpha})$,
identified as the Josephson current. In the following, we shall focus
on the latter while a comprehensive treatment of all terms will be
given in Ref. \cite{future}. Consider now the case of zero externally
applied voltage ($eV=0$).  From Eq. (\ref{eq:currents}), we see that
the Josephson charge-current becomes
\begin{align}
  \label{eq:currents2}
  I^\text{C}_\text{J}(t) =\, & 
  e\sum_{\vk\vp\sigma\alpha} [1+\sigma\alpha\cos\vartheta]\cos(\theta_\vp - \theta_\vk)\sin(\theta^\text{L}_{\sigma\sigma}-\theta^\text{R}_{\alpha\alpha})\notag\\
  &\times|T_{\vk\vp}|^{2}
  |\Delta_{\vk\alpha\alpha}|
  |\Delta_{\vp\sigma\sigma}|F_{\vk\vp}^{\alpha\sigma}/(E_{\vk\alpha}E_{\vp\sigma}),
\end{align}
with $F_{\vk\vp}^{\alpha\sigma} = \sum_\pm [f(\pm
E_{\vk\alpha})-f(E_{\vp\sigma})]/( E_{\vk\alpha} \mp E_{\vp\sigma})$
while the expression for $I^\text{S}_{\text{J},z}(t)$ is equal except
for a factor $(-\alpha/e)$ inside the
summation. Observing that Eq. (\ref{eq:currents2}) may be cast into
the form $I^\mathrm{C}_\text{J} = I_0 + I_m\cos(\vartheta)$, we have
thus found a Josephson current, for both spin and charge, that can be
tuned in a well-defined manner by adjusting the relative orientation
$\vartheta$ of the magnetization vectors. Below, we discuss the
detection of such an effect.

\par 
In the limit where one of the superconducting order parameters
vanishes internally on both sides, \ie the equivalent of an A1-phase,
we see that the interplay between $\vartheta$ and
$\theta_{\sigma\sigma}$ remains, as only one term contributes to the
spin sum over $\{\sigma,\alpha\}$.  In this case, the charge- and
spin-current goes as
$\cos^2(\vartheta/2)\sin\Delta\theta_{\sigma\sigma}$, where
$\Delta\theta_{\sigma\sigma} \equiv
\theta^\text{L}_{\sigma\sigma}-\theta^\text{R}_{\sigma\sigma}$ and
$\Delta_{\vq\sigma\sigma}$ with $\sigma=\{\uparrow,\downarrow\}$,
$\vq=\{\vk,\vp\}$ is the surviving order parameter. For collinear
magnetization $(\vartheta=0)$, an ordinary Josephson effect driven by
the superconducting phase occurs. Interestingly, one is able to tune
this current to zero for $\mathbf{m}_\text{L} \parallel
-\mathbf{m}_\text{R}$ ($\vartheta=\pi$).

\par 
Another result that can be extracted from \eq~(\ref{eq:currents2}) is
a persistent spin-Josephson current even if the magnetizations on each
side of the junction are of equal magnitude and collinear
($\vartheta=0$). This is quite different from the Josephson-like
spin-current recently considered in ferromagnetic metal junctions
\cite{nogueira}. There, a twist in the magnetization across the
junction is required to drive the spin-Josephson effect. In this
Letter, however, we have found a persistent spin-current in the
two-particle channel even for collinear magnetization.
\par
In the special case of $eV=0$ and equal SC phases on each side of the
junction, \ie $\theta_{\sigma\sigma}^\text{L} =
\theta_{\sigma\sigma}^\text{R}$, Eq. (\ref{eq:currents2}) reduces to
the form $I^\text{S}_{\text{J},z} = J_0\sin^2(\vartheta/2)
\sin(\theta_{\downarrow\downarrow}^\text{L}-\theta_{\uparrow\uparrow}^\text{R})$
while $I^\text{C}_\text{J}=0$.  This means that a \textit{two-particle
  spin-current without any charge-current} can arise for non-collinear
magnetizations on each side of the junction in the absence of an
externally applied voltage \textit{and} with equal SC phases
$\theta_{\sigma\sigma}^\text{L}=\theta_{\sigma\sigma}^\text{R}$; see,
however, Ref.  \cite{fnote}.
\par
It is well-known that for tunneling currents flowing in the presence
of a magnetic field that is perpendicular to the tunneling direction,
the resulting flux threading the junction leads to a Fraunhofer-like
variation in the DC Josephson effect, given by a multiplicative factor
$\sin(\pi\Phi/\Phi_0)/(\pi\Phi/\Phi_0)$ in the critical current. Here,
$\Phi_0 = \pi\hbar/e$ is the elementary flux quantum, and $\Phi$ is
the total flux threading the junction due to a magnetic field.
However, this is not an issue in the present case since the
magnetization is assumed to be oriented according to Fig.
\ref{fig:junction}a). Since the motion of the Cooper-pairs is also
restricted by the thin-film structure, there is no orbital effect from
such a magnetization.
\par 
Note that the interplay between ferromagnetism and superconductivity
is manifest in the charge- as well as spin-currents, the former being
readily measurable. Since the critical Josephson currents depend on
the relative magnetization orientation, one is able to tune these
currents in a well-defined manner by varying $\vartheta$. This can be
done by applying an external magnetic field in the plane of the FMSC.
In the presence of a rotating magnetic moment on either side of the
junction, the Josephson currents will thus vary according to Eq.
(\ref{eq:currents2}). Depending on the relative magnitudes of $I_0$
and $I_m$, the sign of the critical current may change. Note that such
a variation of the magnetization vectors must take place in an
adiabatic manner so that the systems can be considered to be in, or
near, equilibrium at all times. Our predictions can thus be verified
by measuring the critical current at $eV=0$ for different angles
$\vartheta$ and compare the results with our theory.  Recently, it has
been reported that a spin-triplet supercurrent, induced by Josephson
tunneling between two $s$-wave superconductors across a ferromagnetic
metallic contact, can be controlled by varying the magnetization of
the ferromagnetic contact \cite{Nature:2006}. Moreover, detection of
induced spin-currents are challenging, although recent studies suggest
feasible methods of measuring such quantities \cite{spinmeasure}.
Observation of macroscopic spin-currents in superconductors may also
be possible via angle resolved photo-emission experiments with
circularly polarized photons \cite{simon2002}, or in spin-resolved
neutron scattering experiments \cite{hirsch}.

\par 
We briefly mention our results in the single-particle channel where we
find that the charge-current and the $\hat{\mathbf{z}}$-component of
the spin-current both vanish for $eV=0$; see Ref. \cite{future} for
details. They are nonzero for $eV\neq 0$ even if the magnetization
vectors are collinear. We stress that the finding of a non-persistent
$\hat{\mathbf{z}}$-component of the spin-current does not conflict
with the results of Ref.  \cite{nogueira}, as their
$\hat{\mathbf{z}}$-direction corresponds to a vector in the $xy$-plane
in our system. For $\Delta_{\vk\sigma\sigma}\to 0$,
$\ve{I}^\text{S}_{\text{sp}}(t) =
2\sum_{\vk\vp}\sum_{\alpha\beta\sigma}
\wdf{\sigma\alpha}\wdf{\sigma\beta}\abs{\Tkp{}}^2
\Imag\{\pauli[\beta\alpha]\Lambda_{\beta\sigma}^{1,1}(-eV)\}$, and the
component of the spin-current parallel to
$\ve{m}_\text{L}\times\ve{m}_\text{R}$ is seen to vanish for
$\vartheta=\{0,\pi\}$ at $eV=0$ in agreement with Ref.
\cite{nogueira}.
\par
We reemphasize that the above ideas should be experimentally
realizable by \eg utilizing various geometries in order to vary the
demagnetization fields. One may also use exchange biasing to an
anti-ferromagnet to achieve non-collinearity \cite{bass1999}. We have
found an interplay between FM and SC in the Josephson channel for
charge- and spin-currents when considering non-unitary spin-triplet
ESP FMSC with coexisting and uniform ferromagnetic and superconducting
order.

\par
\begin{acknowledgments}
  \textit{Acknowledgments}. The authors thank A. Brataas, K.
  B{\o}rkje, and E. K. Dahl for helpful discussions.  This work was
  supported by the Norwegian Research Council Grants No. 157798/432
  and No. 158547/431 (NANOMAT), and Grant No. 167498/V30 (STORFORSK).
\end{acknowledgments}


\begin{thebibliography}{99}  
\bibitem{saxena} S. S. Saxena {\it et. al.}, Nature {\bf 406}, 587 (2000). 
\bibitem{pfleiderer} C. Pfleiderer {\it et. al.}, Nature {\bf 412}, 58 (2001).
\bibitem{aoki} D. Aoki, Nature {\bf 413}, 613 (2001).
\bibitem{kulic} M. L. Kulic, cond-mat/0508276 (2005).
\bibitem{fflo} P. Fulde and R. A. Ferrel, Phys. Rev. \textbf{135}, A550 (1964); A. I. Larkin and Yu. N. Ovchinnikov, Zh. Eksp. Teor. Fiz. \textbf{47}, 1136 (1964).
\bibitem{maeno} Y. Maeno {\it et. al.}, Nature {\bf 372}, 532 (1994).
\bibitem{ishida}   K. Ishida {\it et. al.}, Nature {\bf 396}, 658 (1998).
\bibitem{ber} F. S. Bergeret \etal, Rev. Mod. Phys. \textbf{77}, 1321 (2005).
\bibitem{brataas1} A. Brataas and Y. Tserkovnyak, Phys. Rev. Lett. {\bf 93}, 087201 (2004); Y. Tanaka and S. Kashiwaya, Phys. Rev. B {\bf 70}, 012507 (2004); T. Koyama and M. Tachiki, Phys. Rev. B {\bf 30}, 6463-6479 (1984).
\bibitem{kiselev}  S. I. Kiselev {\it et. al.}, Nature {\bf 425}, 380 (2003).
\bibitem{brataas2} G. E. W. Bauer \etal,  App. Phys. Lett {\bf 82}, 3928 (2003).
\bibitem{brataas3} A. Brataas, \etal, Phys. Rev. B {\bf 66}, 060404 (2002).
\bibitem{leggett1975} A. J. Leggett, Rev. Mod. Phys. \textbf{47}, 331 (1975); Y. M. Bunkov, in \textit{Progress in Low Temperature Physics}, edited by W. P. Halperin (Elsevier, 1995), Vol. XIV, pp. 69.
\bibitem{asanov} G. Rashedi \etal, cond-mat/0501211; Y. Asano, Phys. Rev. B \textbf{72}, 092508 (2005).
\bibitem{tallon1999} J. Tallon \etal, IEEE, Trans. Appl. Supercond. \textbf{9}, 1696 (1999).
\bibitem{kirkpatrick} T. R. Kirkpatrick and D. Belitz, Phys. Rev. Lett. 92, 037001 (2004).
\bibitem{machida}K. Machida and T. Ohmi, Phys. Rev. Lett. \textbf{86}, 850 (2001).
\bibitem{bedell} H. P. Dahal \etal, Phys. Rev. B \textbf{72}, 172506 (2005). 
\bibitem{rudd_pickett_1998} R. E. Rudd and W. E. Pickett, J. Phys. Chem Solids \textbf{59}, 2074 (1998).
\bibitem{huxley} F. Hardy and A. D. Huxley,  Phys. Rev. Lett. {\bf 94}, 247006 (2005).
\bibitem{samokhin} K. V. Samokhin and M. B. Walker, Phys. Rev. B \textbf{66}, 174501 (2002)
\bibitem{tewari} S. Tewari \etal, Phys. Rev. Lett. \textbf{93}, 177002 (2004).
\bibitem{shopova} D. V. Shopova and D. I. Uzunov, Phys. Rev. B \textbf{72}, 024531 (2005).
\bibitem{mineev2} V. P. Mineev and K. V. Samokhin, \textit{Introduction to Unconventional Superconductivity}, Gordon and Breach, New York (1999).
\bibitem{kotegawa} H. Kotegawa \etal, J. Phys. Soc. Jpn. \textbf{74}, 705-711 (2005).
\bibitem{bernhard} C. Bernhard \etal, Phys. Rev. B \textbf{61}, R14960-R14963 (2000).
\bibitem{hu} C. R. Hu, Phys. Rev. Lett. \textbf{72}, 1526 (1994).
\bibitem{tanuma} Y. Tanuma \etal, Phys. Rev. B \textbf{64}, 214519 (2001). See also V. Ambegaokar \etal, Phys. Rev. A {\bf 9}, 2676, (1974); L. J. Buchholtz and G. Zwicknagl, Phys. Rev. B {\bf 23}, 5788 (1981). 
\bibitem{nogueira} F. S. Nogueira and K. H. Benneman,  Europhys. Lett. {\bf 67}, 620 (2004); I. Eremin, F. S. Nogueira and R.-J. Tarento, Phys. Rev. B {\bf 73}, 054507 (2006).
\bibitem{wigner} E. P. Wigner, \textit{Gruppentheorie und ihre Anwendung auf die Quantenmechanik der Atomspektren} Frieder. Vieweg, Braunschweig, 1931). 
$\wdf{\uparrow\uparrow}=\wdf{\downarrow\downarrow}=\cos(\vartheta/2),\wdf{\uparrow\downarrow}=-\wdf{\downarrow\uparrow}=-\sin(\vartheta/2)$; a spin-rotated fermion operator reads $\tilde{d}_{\vp\sigma} = \sum_{\sigma'}\wdf{\sigma'\sigma}d_{\vp\sigma'}$.
\bibitem{shi2} J. Shi \etal, Phys. Rev. Lett. 96, 076604 (2006).
\bibitem{future} J. Linder, M. Gr{\o}nsleth, A. Sudb\o ,  
submitted to Phys. Rev. B, {\tt cond-mat/0609314}.
\bibitem{Nature:2006} R. S. Keizer \etal,
Nature {\bf 439}, 825 (2006).
\bibitem{spinmeasure} A. G. Mal'shukov \etal,
Phys. Rev. Lett. {\bf 95}, 107203 (2005).
\bibitem{simon2002} M. E. Simon and C. M. Varma, Phys. Rev. Lett. \textbf{89}, 247003 (2002).
\bibitem{hirsch} J. E. Hirsch, Phys. Rev. Lett {\bf 83}, 1834 (1999).
\bibitem{bass1999} J. Bass and W. P. Pratt Jr., Journal of Magnetism and Magnetic Materials \textbf{200}, 274 (1999).
\bibitem{fnote} This effect requires that
$\theta_{\uparrow\uparrow}^\text{L(R)}-\theta_{\downarrow\downarrow}^\text{L(R)}
  \neq \{0,\pi\}$. If an interband Josephson coupling is present, the
  phases will be locked according to
  $\theta_{\uparrow\uparrow}-\theta_{\downarrow\downarrow} =
  \{0,\pi\}$. It has recently [J. Shi and Q. Niu, cond-mat/0601531 (2006)] been proposed that $p$-wave
  SC could be induced by means of spin-orbit coupling in FM metals,
  which in turn leads to an interband Josephson coupling.  However, in
  the general case Eq. (\ref{eq:currents2}) is non-zero even if the
  spin-up and spin-down phases are locked to each other.
\end{thebibliography}
\end{document}